\documentclass{aa}
\usepackage{graphicx}
\usepackage{natbib}
\bibpunct{(}{)}{;}{a}{,}{,}

\newcommand{\td}{T_{90}}
\newcommand{\tem}{{\cal T}_{50}}

\newcommand{\fl}{\cal F}
\newcommand{\Epk}{E_{\rm pk}}
\newcommand{\Epkin}{E_{\rm pk;intr}}
\newcommand{\REpk}{{\cal R}E_{\rm pk}}
\newcommand{\SF}{{\cal S}_{\cal F}}
\newcommand{\lag}{\tau_{\rm lag}}

\begin{document}

\title{Factor analysis of the long gamma-ray bursts}

\titlerunning{Factor analysis of the long gamma-ray bursts}

\author{Z. Bagoly \inst{1}
\and
L. Borgonovo \inst{2}
\and
A. M\'esz\'aros \inst{2,3}
\and
L. G. Bal\'azs \inst{4}
\and
I. Horv\'ath \inst{5}
}

\offprints{Z. Bagoly}

\institute{
Dept. of Physics of Complex Systems, E\" otv\" os
          University, H-1117 Budapest, P\'azm\'any P. s. 1/A,
          Hungary\\
          \email{zsolt.bagoly@elte.hu}
       \and
Stockholm Observatory, AlbaNova, SE-106 91 Stockholm, Sweden\\
 \email{luis@astro.su.se}
\and
               Astronomical Institute of the Charles University,
              V Hole\v{s}ovi\v{c}k\'ach 2, CZ 180 00 Prague 8,
          Czech Republic\\
              \email{meszaros@mbox.cesnet.cz}
\and
       Konkoly Observatory, H-1525 Budapest, POB 67, Hungary\\
              \email{balazs@konkoly.hu}
\and
              Department of Physics, Bolyai Military
                    University, H-1581 Budapest, POB 15, Hungary\\
              \email{horvath.istvan@zmne.hu}
              }

   \date{Received September 7, 2007; accepted  ..........}

\abstract
{}
{We study statistically 197 long gamma-ray bursts, detected and measured in detail by the
BATSE instrument of the Compton Gamma-Ray Observatory. In
the sample 10 variables, describing for any burst the time behavior of the
spectra and other quantities, are collected.}
{The factor analysis method is used to find the latent random 
variables describing the temporal and spectral properties of GRBs.}
{The application of this particular method to this sample indicates that five
factors and the 
$\REpk$ spectral variable (the ratio of peak energies in the spectrum)
describe the sample satisfactorily. Both the pseudo-redshifts
inferred from the variability, and the Amati-relation  in its original
form, are disfavored.}
{}


\maketitle

\section{Introduction}

	Factor Analysis (FA) and Principal Component Analysis (PCA) are
powerful statistical methods in data analysis.  
Using PCA and FA \cite{bag98} demonstrated that 
the 9 variables typically measured ($T_{50}$ and $T_{90}$ durations; $P_{64},
P_{256}$, and $P_{1024}$ peak fluxes; ${\cal F}_1, {\cal F}_2, {\cal F}_3$, and
${\cal F}_4$ fluences) for gamma-ray bursts (GRBs), observed by the BATSE
instrument onboard the Compton Gamma-Ray Observatory and listed in the Current
BATSE Catalog \citep{mee01}, can be satisfactorily represented by 3 hidden
statistical variables. \cite{borbjor06} (hereafter BB06) studied the
statistical properties of 197 long GRBs detected by BATSE. They defined 10
statistical variables describing the temporal and spectral properties of GRBs.
By performing a PCA, they concluded that about 70~\% of the total variance in the
parameters were explained by the first 3 Principal Components (PCs).  The aim
of this article is to proceed in a similar way to BB06 by using instead FA.

	By solving the eigenvalue problem of the correlation (covariance)
matrix, PCA transforms the observed variables into the same number of uncorrelated
variables (PCs).  An essential ingredient of PCA is a
distinction between the ``important'' and ``less important'' variables by taking
into account  the magnitude of the eigenvalues of the correlation (covariance)
matrix.  FA assumes that the observed variables can be described by a linear
combination of hidden variables given by:

\begin{equation}
\label{eq1}
{\bf x = \bf \Lambda f + \varepsilon}\,,
\end{equation}

\noindent where ${\bf x}$ denotes  an observed variable of dimension $p$, ${\bf
\Lambda}$ is a matrix of $p \times m$ dimensions ($m<p$), $f$ represents a hidden
variable of $m$ dimensions. The components of ${\bf \Lambda}$ are called
loadings, the factor $f$ represents scores, and  $\varepsilon$ is a noise term.
We can infer ${\bf x}$ from observations while the quantities on the right-hand-side of
Eq.~\ref{eq1} have to be computed by a suitable algorithm.

	PCA expresses the ${\bf x}$ observed variable as a linear
transformation of a hidden variable of the same $p$ dimension, whose components
are uncorrelated.  The transformation matrix is set up from the eigenvectors of
the correlation matrix of ${\bf x}$. By retaining only the first $m<p$ eigenvectors, 
it can be shown that, 
the resultant transformation matrix provides the best
reproduction of ${\bf x}$ among those using only $m < p$ components. By retaining 
only the first $m < p$ eigenvectors, one receives a transformation matrix of dimensions 
$p \times m$ and an expression identical to the  first term on the right
side of Eq.~\ref{eq1}. Due to this fact, the PCA is a default solution of FA
in many statistical packages (e.g. SPSS\footnote{SPSS is a registered
trademark (see {\em www.spss.com})}; for a detailed comparison of PCA and FA,
see \cite{jo02}).  Although PCA is a default solution in many packages, FA has
other algorithms as well. In our computations, we use the Maximum Likelihood
(ML) method (for details see \cite{jo02}).

\section{The sample}

	We use the sample of 197 long  GRBs in BB06 and the 10 variables
defined there.  Of the 10  variables, $\td$ and  $\fl$ were 
taken directly from the BATSE Catalog.  The remaining 8 variables were calculated by BB06.
In summary, the 10 variables are the following: duration time $\td$, emission
time $\tem$, autocorrelation function (ACF) half-width $\tau$, variability $V$,
emission symmetry $\SF$, cross-correlation function time lag $\lag$, the ratio
of peak energies $\REpk$, fluence $\fl$, peak energy $\Epk$, and low frequency
spectral index $\alpha$.

\begin{table*} \caption{Correlation matrix among the 10 variables. Except for
$\alpha$ the decimal logarithms are taken.} $$
\begin{array}{rrrrrrrrrrrr} \hline 
Variable & \log \td & \log \tem & \log \tau & \log V  & \log \SF & \log \lag & \log \REpk  &  \log \fl &  \log \Epk & \alpha \\
\hline
\log \td & 1.00 & 0.78 & 0.58 & 0.18 & 0.09 & -0.01 & -0.15 & 0.5& 0.24 &    -0.26 \cr
\log \tem & 0.78 & 1.00 & 0.87 & 0.51 & 0.25& 0.09 & -0.21 & 0.61 & 0.14 &  -0.16 \cr
\log \tau & 0.58 & 0.87 & 1.00 & 0.4 & 0.24 & 0.15 & -0.25 &  0.61 & 0.14 &    -0.12 \cr
\log V & 0.18 & 0.51 & 0.4 & 1.00 & 0.32 & -0.18 & -0.37 &  0.33 & 0.08 &    -0.07 \cr
\log \SF & 0.09 & 0.25 & 0.24 & 0.32 & 1.00 & 0.03 & -0.37 &  0.07 & -0.23 & 0.03 \cr
\log \lag & -0.01 & 0.09 & 0.15 & -0.18 & 0.03 & 1.00 & 0.24 & -0.04 &      -0.28 & 0.33 \cr
\log \REpk &  -0.15 & -0.21 & -0.25 & -0.37 & -0.37 & 0.24 & 1.00 &  -0.03 & 0.04 & -0.01 \cr
\log \fl &      0.5 & 0.61 &  0.61 & 0.33 & 0.07 & -0.04 & -0.03 & 1.00 & 0.58 & -0.2 \cr
\log \Epk & 0.24 & 0.14 & 0.14 & 0.08 & -0.23 & -0.28 & 0.04 & 0.58 & 1.00 & -0.28 \cr
\alpha & -0.26 & -0.16 & -0.12 & -0.07 & -0.03 & 0.33 & -0.01 & -0.2 & -0.28 & 1.00 \cr
\hline \end{array} $$
\label{correlation}
\end{table*}

	Since the variables have different dimensions in a similar way to BB06 we use
the decimal logarithms (except for $\alpha$). The correlations between the
variables are indicated in Table~\ref{correlation}.  The choice of the logarithms
is motivated by the fact that the distributions of most variables are well
described by log-normal distributions (see the discussion of BB06). 

	In a similar way to BB06, we do not consider
the fluence on the highest channel ($> 300$ keV) separately, although in
\cite{bag98} this variable alone was used to define a PC (factor).  This choice is
motivated by two reasons: first, fluences on the fourth
channel often vanish or have significant errors (``the values are noisy'');
second, as noted by BB06, in a sample of long-soft GRBs {\it only},
this quantity is less important. It is now certain that the long-soft and
short-hard  bursts are different phenomena \citep{ho98,nor01,ho02,bal03}. The
significance of the intermediate GRBs is unclear \citep{ho06}.

\section{Estimation of the number of factors}

	In contrast to PCA, in FA the choice of the number of hypothetical
(latent) random variables (factors) is - at the beginning - a free parameter.  To determine
the optimal number of factors, there are no direct methods (even the notion ``best
number of factors'' is unclear; see \cite{jo02}).

	By solving the eigenvalue problem of the correlation matrix, PCA yields
PCs in descending order of the eigenvalue magnitudes.  To validate a
factor model, one retains the first $m < p$ PCs, which satisfactorily reproduce
the original correlation matrix. In the ML method, the expected number of
factors is an input parameter, and the algorithm computes the probability that
the difference between the original and reproduced correlation matrix can be attributed to
chance only.  One stops increasing the number of factors, when this probability
is already sufficiently large.

	The factor model assumes that a linear transformation exists between the observed
and the latent (factor) variables. The number of unknown parameters (i.e. $p\:(m+1)$
on the right side of Eq.~\ref{eq1}) are constrained by the dimension of the
covariance matrix of {\bf x} (i.e. $1/2\:p(p+1)$ independent parameters) and the
need for factor-loading orthogonality, which provides $1/2\:m(m-1)$ free parameters
(\cite{KS73}).  Thus, the number $m$ of factors can be constrained by the
following inequality: 
\begin{equation} 
\label{constr} 
m \leq (2p+1- \sqrt{8p+1})/2 \:, 
\end{equation} 
\noindent which provides $m \leq 6 $ in our case.  Since the number of factors is
an integer, $m=6$ is a maximum value in our case.  Equation \ref{constr} provide the 
upper limit to the number of factors, although the true number remains to
be estimated.

	There are several further criteria that constrains the required number of
factors (\cite{jo02} and references therein).  The first additional criterion
follows from the ``cumulative percentage of the total variance.'' Taking into
account any new factor, the percentage of the variation explained by these
factors should increase. Then, if one defines a cut-off percentage, the number
of factors $m$ required is given by the value factors, when the cumulative
variance in percentage is already higher than this cut-off percentage. There is
no exact rule about the best value of the cut-off: \cite{jo02} %Jolliffe (2002) (Chapt. 6.1.1.)
proposes to choose a value around 70\% - 90\%, and in addition, if $p >> 1\%$, a
smaller value is proposed. Hence, in our case the value around 70\% seems to be
a good choice.  For PCA and for the correlation matrix, $m$ can also be estimated
from the eigenvalues of the PCs - PCs with eigenvalues larger than 0.7
should be retained. Using FA - instead of the PCA - one may also assume that the
number of factors in general should not be larger than the number of PCs (in most
cases it is even smaller) \citep{jo02}.
The most accurate estimate of the number of factors m is therefore a combination of several criteria.

	The advantage of the ML approach is that it helps to constrain the value of
$m$, the dimension of the hidden factor variables.  This is because
the ML method provides a probability of the null hypo\-thesis, i.e. that the
correlation matrix of the observed variables and that reproduced by the factor
solution are identical from the statistical point of view.

\begin{table*}
\centering
\caption{
ML solution assuming 6 factors. 
In any column for the given factor the loadings are given (a larger value represents
higher weight for a given variable); the sum of their squares is denoted by
{\em SS loading}; the value {\em Proportion Var} defines the proportion of {\em
SS loading} to the sum of variances of the input variables; {\em Cumulative
Var} defines the sum of proportional variances.
} \label{fac6}
\begin{tabular}{lrrrrrr}
\hline {\em Variable}              & {\em Factor1} & {\em Factor2} & {\em Factor3} & {\em Factor4} & {\em Factor5} & {\em Factor6}\\
\hline
$\log\td$       &   0.418  & 0.128 & -0.066 &  0.884 & -0.133 &  0.017 \\
$\log\tem$     &   0.770  & 0.022 & -0.087 &  0.490 & -0.036 &  0.320 \\
$\log\tau$       &   0.928  & 0.038 & -0.158 &  0.198 & -0.006 &  0.146 \\
$\log V$    & 0.249  & 0.063 & -0.225 &  0.043 & -0.041 &  0.844 \\
$\log\SF$     &   0.173 & -0.241 & -0.319 &  0.036 & -0.042 &  0.252 \\
$\log\lag$       &   0.246 & -0.269 &  0.235 & -0.008 &  0.333 & -0.187 \\
$\log\REpk$    &  -0.070 &  0.001 &  0.981 & -0.050 &  0.003 & -0.159 \\
$\log\fl$   &   0.564 &  0.499 &  0.047 &  0.226 & -0.066 &  0.187 \\
$\log\Epk$    &  0.108  &  0.974 &  0.054 &  0.074 & -0.159 & -0.008 \\
$\alpha$          &  -0.098 & -0.105 & -0.024 & -0.106 &  0.981 & -0.004 \\
\hline
{\em SS~loadings}   &   2.126 &  1.363  & 1.212   & 1.134   & 1.126   & 0.995 \\
{\em Proportion~Var}&   0.213 &  0.136  & 0.121   & 0.113   & 0.113   & 0.099 \\
{\em Cumulative~Var}&   0.213 &  0.349  & 0.470   & 0.584   & 0.696   & 0.796 \\
\hline
\end{tabular}
\end{table*}

	By performing FA on the observed variables assuming 6 factors, which is the maximum
number allowed by Eq.~\ref{constr}, one observes the validity of the null
hypotheses with only $p=0.0191$, which implies that even the maximum allowable number
of factors can't reproduce the original correlation matrix of the observed
variables satisfactorily. Table \ref{fac6} shows the factor coefficients
(loadings) of this solution.

	By inspecting Table \ref{fac6}, it becomes obvious that {\em Factor3} and
{\em Factor5} are dominated by only one variable ($\log \REpk$ and $\alpha$,
respectively) and are hardly affected by the other variables.
Therefore,
it appears reasonable to exclude one of them and repeat the calculations with the
remaining 9 variables. In this case, the maximum allowable number of factors is
$m=5$, which corresponds to either the null hypotheses $p=0.11$, after
excluding $\alpha$, and $p=0.273$ after excluding $\log \REpk$. We 
therefore decided to exclude $\log \REpk$, and the ML solution assuming $m=5$ factors
is given in Table \ref{fac5}.  The cumulative variance, defined by 5 factors,
is $71.9\%$. This fulfills the ``cumulative percentage of the total variance''
criterion for PCA, considering the corresponding high value of $p$. This also
supports the choice of 5 factors.

	We have proven that $m=5$ factors are sufficient.  To prove that it is
essential, we also performed the ML analysis with $m=4$ factors. This
calculation resulted only $p=0.0044$ that 4 factors are sufficient. One can
therefore conclude that $m=5$ factors are necessary and sufficient for
describing the observed variables.

\begin{table*}
\centering
\caption{
ML solution assuming 5 factors after removing the $\log \REpk$ variable.
Testing  the hypothesis that 5 factors are sufficient resulted $p=0.273$.
}\label{fac5}

\begin{tabular}{lrrrrr}
\hline {\em Variable} & {\em Factor1} & {\em Factor2} 	& {\em Factor3} 	& {\em Factor4} 	&   {\em Factor5}
\\\hline
$\log\td$	&	0.875	&	0.009	&	0.088	&	-0.152	&	-0.051	\\
$\log\tem$	&	0.895	&	0.353	&	0.039	&	0.026	&	0.236	\\
$\log\tau$	&	0.704	&	0.277	&	0.090	&	0.095	&	0.592	\\
$\log V$	&	0.176	&	0.973	&	0.091	&	-0.098	&	0.016	\\
$\log\SF$	&	0.133	&	0.320	&	-0.244	&	-0.020	&	0.141	\\
$\log\lag$	&	0.110	&	-0.144	&	-0.175	&	0.490	&	0.141	\\
$\log\fl$	&	0.528	&	0.183	&	0.520	&	-0.068	&	0.245	\\
$\log\Epk$	&	0.146	&	-0.060	&	0.947	&	-0.272	&	-0.005	\\
$\alpha	$	&	-0.191	&	0.038	&	-0.053	&	0.730	&	-0.100	\\
\hline
{\em SS~loadings}   &   2.459 &  1.309  & 1.285  & 0.895  &  0.519 \\
{\em Proportion~Var}&   0.273 &  0.145  & 0.143  & 0.099  &  0.058 \\
{\em Cumulative~Var}&   0.273 &  0.419  & 0.561  & 0.661  &  0.719
\\\hline
  \end{tabular}
\end{table*}

\section{Results and discussion of FA}

	The first factor is constrained by $\td,\, \tem,\, \tau\,$ and $\fl$, i.e.
the first factor is determined mainly by the temporal properties.  Hence measures
$\tem$ and $\td$ are the preferred length indicators over $\tau$.

	The second factor is dominated by $V$.  However, according to
\cite{rafe00}, \cite{re01}, and \cite{2005MNRAS.363..315G}, the variability
should be correlated with the luminosities of GRBs, and hence to the fluence.
No significant connection is, however, inferred by the second factor raising
queries about the redshift estimations derived from variability.

	The third factor is mainly driven by $\Epk$.  It is interesting that
the peak energy in the spectra appears to dominate the third factor
so significantly. It emphasizes that the spectrum itself is an
important quantity (an expected result), and, in the spectrum  $\Epk$
itself, is a significant descriptor (an unexpected result). 
In addition, the loading of $\fl$ is also important to the 
third factor. All this has a remarkable impact on the Amati-relation. 

	The Amati-relation (\cite{am02}) proposes that there should be a linear
connection between $\log \Epkin$ and $\log E_{iso}$, where $E_{iso}$ is the
emitted energy under the assumption of isotropic emission, $\Epkin$ $= (1+z)
\Epk$ is the intrinsic peak energy, and $z$ is the redshift.  This relation, which
follows from the relation $\Epkin \propto E_{iso}^{x}$ found by \cite{am02}
from the analysis of twelve bright long GRBs with well-measured
redshifts. The most probable value of $x$ was around $x = 0.5$. Thus, the
Amati-relation - in its original form - claims that a direct linear connection exists {\it
only} between $\log \Epkin$ and $\log E_{iso}$.  We note that the Amati-relation was
predicted even earlier by the strong correlation between $\log \fl$ and $\log
\Epk$ \citep{Lloyd00}.  The importance of the Amati-relation is
straightforward: if it  holds, then it is possible to determine the redshift of
the given long burst from the value of  $\Epk$ {\it alone}, because  $\Epk$
defines   $E_{iso}$ independently of $\fl$. Then, by applying standard cosmology, we can 
calculate from the known $E_{iso}$ and $\fl$ values the redshift (e.g.  \cite{mm95}). 

	The validity of the Amati-relation has been a matter of intense
discussion since publication.  Several papers confirmed it by newer
analyses (e.g.  \cite{am06,ghir07, ghir08} and references therein).  
\cite{cab07} confirmed the existence of the $\Epkin$ - $E_{iso}$ correlation in the
rest-frame for 47 Swift GRBs.
These studies considered bright long GRBs with known redshifts enabling $E_{iso}$ to be determined.  
This causes strong selection effect in the
studied samples. 
It is possible that this selection effect cause e.g. the entire BATSE sample to
follow the Amati-relation either only in a modified version or even not at all,
even though the relation holds for the truncated sample of bright GRBs
\citep{napi05,but07}.  BB06 obtained that it is better to use
$\Epkin \propto E_{iso}^{a_1}\tau_{intr}^{b_1}$ with suitable $a_1$ and $b_1$
for the BATSE sample ($\tau_{intr} =\tau/(1+z)$).  Hence, if $b_1 \neq 0$, then
the Amati-relation is altered. BB06 proposes, as the optimal choice,
$b_1 = -0.3$.  Some papers even reject the Amati-relation both in
the BATSE sample \citep{napi05} and in the Swift sample \citep{but07}. The
most radical solution even challenges the meaning of $\Epkin$ itself in the
spectra of GRBs \citep{ry05b}.

	For our purposes, it is essential statistically that the correlation between $\log \fl$ and
$\log \Epk$ does not imply that there is a linear connection {\it only} between
$\log E_{iso}$ and $\log \Epkin$. BB06 also arrived at the
conclusion that a relation of the form 
\begin{equation} 
\log E_{iso} = a_1 \log \Epkin  + b_1 \log \tau_{intr} + c_1 
\end{equation} 
should exist with some suitable non-zero constants $a_1, b_1,$ and $c_1$.  We note that
$\tem$ and $\tau$ strongly correlates with each other, i.e. in this equation
either $\tau_{intr}$ or ${\cal T}_{50;intr}$  can be used.

	The factor loadings imply that $\log \fl$ is explained basically by the
first and third factors.  Since in {\em Factor1} and {\em Factor3} 
 $\log \tau$ and $\log \Epk$ are very strong, respectively, it suggests that
\begin{equation}
\label{epkinmod}
\log E_{iso} = a_2 \log \Epkin + b_2 \log \tau_{intr} + c_2 \log L_{iso} +d
\end{equation}
should hold with some suitable $a_2, b_2, c_2,$ and $d$ non-zero constants ($L_{iso}$
is the isotropic peak luminosity).  We note that a similar relation was also proposed
by \cite{Fir06}.

	The correlation between $\log \fl$ and $\log \Epk$ is mainly
determined by {\em Factor3}.  It follows from the loadings of the first and
third factors that the relationship between $\log \fl$ and $\log \Epk$ is as
important as with the variables dominating {\em Factor1}.  This fact disfavors
a simple linear relationship {\it only} between  $\log \Epkin$ and $\log
E_{iso}$. The detailed study of Eq.~\ref{epkinmod} (cf.  determination of $a_2,
b_2, c_2, d$, and alternative equations) is beyond the aim of this paper.
Even from this conclusion, it however follows that the 
Amati-relation in its original form is disfavored and some modified 
version proposed by BB06 is also supported here.

	The fourth factor is defined by low frequency spectral index $\alpha$
and $\lag$.  This implies that the direct correlation between $\lag$ and $V$ is
negligible, and hence there is no direct support for the luminosity estimators
based on these two variables \citep{rafe00,re01,nor02}.

	The fifth factor is dominated by $\tau$ and $\fl$. 
With the first factor this demonstrates that  $\td$ and $\tem$ are not completely
equivalent, although $\tem$ characterizes a burst more closely.

	In our opinion, the most remarkable result is that so few quantities are
needed, i.e. that all nine quantities can be characterized by five
variables.  Because all of these conclusions are derived from the measured data
alone, all models of GRBs must respect these expectations.

	The number of essential variables is in accordance with BB06. They
claimed that 3-5 PCs should be used, and we constrained the number of important
quantities to be 5. % corresponding to 4 PCs in BB06.

\section{Conclusions}

The results of the paper may be summarized as follows.

\begin{itemize}

\item No more than 5 factors should be introduced. This essential lowering of
the significant variables is the key result of this paper.

\item The structure of factors is similar to the PCs of BB06.  The number of
important quantities is more accurately defined here.

\item The first factor is dependent mainly on the temporal variables, and 
quantities $\tem$ and $\td$ are the preferred length indicators.

\item The second factor is dominated by the variability.

\item The connection of $\Epk$ in the third factor with other quantities, and
the structure of the first three factors cast some doubts about the
Amati-relation in its original form.

\item The $\alpha$ and $\lag$ parameter values in fourth factor give no direct support for the
luminosity estimators.

\item The fifth factor demonstrates that $\td$ and $\tem$ are not completely
equivalent.

\end{itemize}

\begin{acknowledgements}

Thanks are due for the valuable discussions to Claes-Ingvar Bj\"ornsson,
Stefan Larsson, Peter M\'esz\'aros, Felix Ryde, P\'eter Veres, and the anonymous
referee.  This study was supported by the Hungarian OTKA grant No. T48870 and
Bolyai Scholarship (I.H.), by a Research Program MSM0021620860 of the Ministry
of Education of Czech Republic, by a GAUK grant No. 46307, and by a grant from
the Swedish Wenner-Gren Foundations (A.M.).

\end{acknowledgements}

\end{document}